\documentclass[11pt]{article}
\usepackage{hyperref}
\usepackage{moriond,epsfig}
\usepackage{graphicx}

\def\be{\begin{equation}}
\def\ee{\end{equation}}
\def\bea{\begin{eqnarray}}
\def\eea{\end{eqnarray}}

\begin{document}
\vspace*{4cm}
\title{THROUGH A GLASS, DARKLY: THEORY SUMMARY}

\author{ DAVISON E. SOPER }

\address{Institute of Theoretical Science, University of Oregon, \\
Eugene, Oregon 97403, USA}

\maketitle\abstracts{
This is a summary of the theoretical contributions to the QCD session of the 47th Rencontre de Moriond, including some perspectives on the implications of the reported experimental results on the status of our theoretical understanding.
}

\centerline{\em For now we see through a glass, darkly, but then face to face:}
\centerline{\em now I know in part; but then shall I know \dots}

\medskip
\section{Introduction}
\label{sec:intro}
The quote above is from 1 Corinthians 13 in the King James Bible. To my mind, it illustrates our situation at the 47th Rencontre de Moriond ``QCD and High Energy Interactions"~\cite{moriond} as we heard of tantalizing hints from experiment of the existence of the Higgs boson, but wait to know whether these hints will take convincing form or recede into the dark mists in the 2012 running of the CERN Large Hadron Collider. 

The question to be answered is whether we are confirming the Standard Model mechanism for electroweak symmetry breaking or are finding structures that lie beyond the Standard Model. In order to present a definite point of view on this, I take the Standard Model to be a renormalizable quantum field theory, viewed as a low energy effective field theory with a high cutoff energy scale.

With this view, there are good arguments that the Standard Model is wrong. The issues were nicely presented in the talk of G.~Altarelli \cite{altarelli}. If the energy cutoff scale is very large, then it is difficult to understand why the Higgs boson mass not also very large. I will discuss this point when we come to the Higgs boson in Sec.~\ref{sec:higgs}. Also, there is solid astronomical evidence for dark matter particles that are not present in the Standard Model. Additionally, there are some experimental anomalies. For instance, the forward-backward asymmetry in $p \bar p \to t\bar t$ and the proton charge radius as measured from the energy levels of muonic hydrogen do not seem to fit well with Standard Model expectations. I will return to these issues in Sec.~\ref{sec:flavor}.

Despite these indications that the Standard Model is wrong, it has passed many, many experimental tests. I suppose that it seems as durable as the Roman Empire must have seemed to those living in La Thuile two thousand years ago. In contrast, physics beyond the Standard Model offers prospects that seem now unknown and largely unknowable until we find the sought indications from experiment.

My plan for this summary is to say something about each of several topics covered at the conference: emergent phenomena within the Standard Model (Sec.~\ref{sec:emergent}); testing Standard Model physics (Sec.~\ref{sec:SM}); direct searches for new physics (Sec.~\ref{sec:newphysics});  indirect searches for new physics using flavor physics (Sec.~\ref{sec:flavor}); and looking for the Standard Model Higgs boson (Sec.~\ref{sec:higgs}).

\section{Emergent phenomena within the Standard Model}
\label{sec:emergent}

We heard about some kinds of phenomena that are presumed to arise from the Standard Model lagrangian but that do not come about in a simple way, so that highly non-trivial theoretical insights are needed if we are to have a real understanding of the observations. I discuss these in three categories.

\subsection{Transverse flow in heavy ion collisions}
\label{sec:heavyionflow}

In a heavy ion collision, one expects that the flow of energy in the plane transverse to the beam axis will not be symmetric under rotations but rather will depend on the angle between the measured momentum and the direction defined by the transverse vector $\vec b$ from the center of one of the nuclei to the center of the other. (One can measure $\vec b$ approximately from the total activity in the event and the geometry of this activity.) Experimental results on this flow were reported by STAR and PHENIX,\cite{RHICflow} ALICE,\cite{ALICEflow} CMS,\cite{CMSflow} and ATLAS.\cite{ATLASflow} These are an experimental results, but some comments about the theory may be helpful.

\begin{figure}
\centerline{
\includegraphics[width = 15 cm]{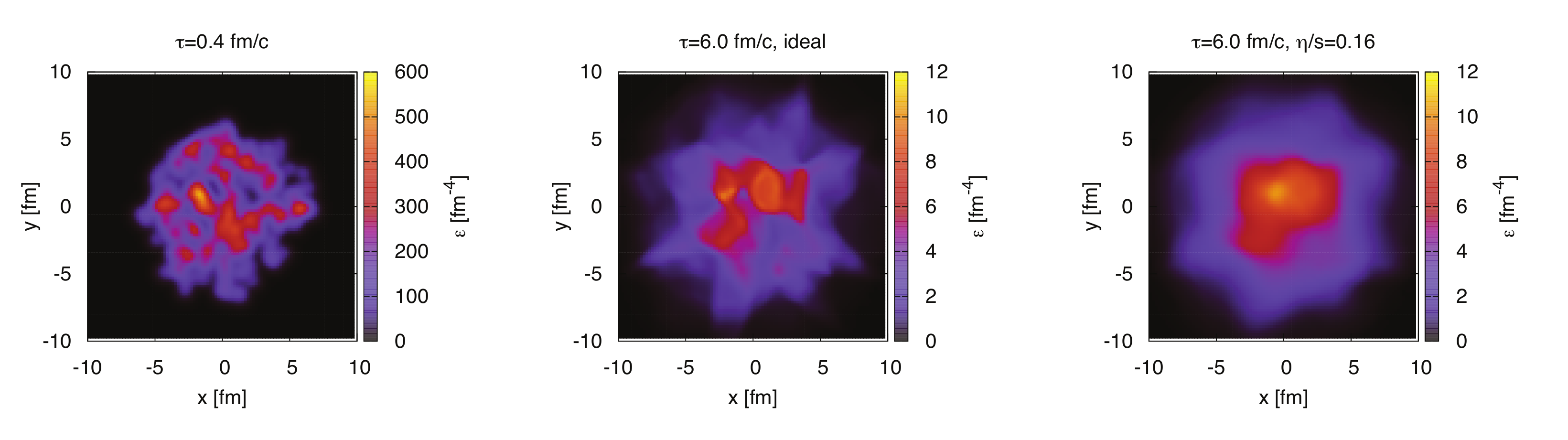}
}
\caption{Modeled~\protect\cite{jeon} energy density distribution in the transverse plane for one event at the initial time (left) and after a time of 6 fm/c for the ideal fluid case (middle) and with $\eta/s = 0.16$ (right), as reported in the talk of R.~Snellings from ALICE.\protect\cite{ALICEflow}.
\label{fig:flow}}
\end{figure}

Suppose that $|\vec b|$ is not small compared to the nuclear radius $R$, but not close to $R$ either. Then the region in the transverse plane in which constituents of the nuclei collide is almond shaped, with the long axis of the almond perpendicular to $\vec b$. Presumably, just after the collision, this region is filled with a hot plasma of some sort, with very high pressure in the middle and low pressure on the outside. (The use of this language implies a system that is locally not too far from thermodynamic equilibrium, but one has to judge from the data how close to reality such a hydrodynamic picture is.) Because the pressure gradient is greatest in the directions parallel or antiparallel to $\vec b$, one expects that the matter in the plasma will gain momentum in these directions. After the plasma has expanded and cooled, one then expects that the transverse particle flow as measured by $dN/d\phi$ will be biggest in the $\pm \vec b$ directions, producing a pattern $N_0[1 + 2 V_2 \cos(2(\phi - \phi_b))]$ with nonzero $V_2$. This is what the experiments find.  There are also contributions proportional to $V_n \cos(n(\phi - \phi_b))$.
Many interesting results along these lines were presented. When interpreted using a model of hydrodynamic flow~\cite{jeon} with a model for the density fluctuations at the initial time, the results suggest that the viscosity is quite small, as illustrated in Fig.~\ref{fig:flow} from the talk of R.~Snellings from ALICE.\cite{ALICEflow}. More generally, the experimental results suggest that the mean free path within the medium is small.

\subsection{Jet quenching in heavy ion collisions}
\label{sec:quenching}

There were also interesting results from ATLAS~\cite{ATLASquenching} and CMS~\cite{CMSquenching} about jet quenching in heavy ion collisions. Here one attempts to investigate back-to-back jets produced by parton-parton collisions with a transverse momentum given to the jets. If the partonic collision happens near the edge of the transverse collision region and if jet A is produced heading toward the edge of the hadronic matter while jet B is produced heading toward the middle, then it seems clear that jet B should suffer a substantial energy loss in the medium. This effect is seen in the data for high transverse momentum jets. The size of the effect and its dependence on the available event parameters should be able to test our understanding of the underlying physics. As a parton is passing through the nuclear matter, it scatters from the partons in the nuclear matter and emits bremsstrahlung gluons, which further scatter. Thus a jet develops quite differently from a jet that creates a parton shower in vacuum. It seems to me that developing a detailed picture of this, a picture that can match the data, will be a significant challenge to theorists. 

\subsection{Physics of the saturation scale}
\label{sec:saturation}

In QCD at small momentum fraction $x$, the saturation scale $Q_s(x)$ is a key concept. The saturation scale can be qualitatively defined by the relation $\alpha_s x f_g(x,Q_s^2)/[Q_s^2 R_p^2] = 1$. Here $\alpha_s/Q^2$ is the cross section for a gluon in a proton to scatter from another parton with a momentum transfer of scale $Q$. Since $x f_g(x,Q^2)$ is the number of gluons per unit $\log(x)$, the product $\alpha_s x f_g(x,Q^2)/Q^2$ is the transverse area in the proton covered by gluons. The transverse area of the proton is proportional to the square of its radius, $R_p$. Thus $\alpha_s x f_g(x,Q^2)/[Q^2 R_p^2]$ measures the fraction of the proton's area covered by gluons. That is, for a given $x$, the proton appears black for scattering processes of scale $Q_s$ and below. For small $x$, $x f_g(x,Q_s^2)$ is big enough that $Q_s(x)$ becomes greater than a GeV, so that the physics of the saturation scale is at least marginally perturbative. 

We did not have a session devoted to physics of the saturation scale, but the idea was present in several talks. See, for instance, the talk of M.~Perdekamp~\cite{perdekamp} on the suppression of hadrons at forward rapidity at RHIC.

M.~Praszalowicz~\cite{praszalowicz} proposed a scaling formula involving $Q_s(x)$ in which the $p_T$ distribution of produced particles in hadron-hadron collisions becomes a function of one variable instead of the two variables $p_T^2$ and $s$.

I.~Sarcevic~\cite{sarcevic} showed a calculation of the rate at which neutrinos are produced from the decay of charm produced in collisions of cosmic rays with air nuclei. Here $x$ is very small, so she used a color dipole model that that incorporates $Q_s(x)$. The same color dipole/saturation model appeared in the talk of M.~Sadzikowski.\cite{sadzikowski} The issue here is diffractive deeply inelastic electron scattering. This is governed by diffractive parton distribution functions that obey the DGLAP evolution equation at large $Q^2$, but at small $Q^2$ and very small momentum fraction, saturation effects take over.

\subsection{Use of gauge-gravity duality}
\label{sec:AdsCFT}

M.~Djuric~\cite{djuric} reported on studies of the pomeron using the conjectured connection between field theory at large coupling and higher dimensional gravity at weak coupling (the AdS/CFT correspondence). Specifically, he analyzed deeply virtual Compton scattering from this point of view. R.~Brower~\cite{brower} reported on studies of diffractive Higgs production using this same picture.

\section{Calculating and testing Standard Model physics}
\label{sec:SM}

Calculations of Standard Model cross sections continue to improve. This is important because Standard Model processes are important backgrounds for many new physics signals that we are looking for. The better we know the background, the better we can find the signal. In addition, the same calculational techniques allow us to better calculate cross sections for possible new physics signals, particularly when the sought new particles carry color. Finally, we can compare calculated Standard Model cross sections to data. This tests our ability to calculate correctly and to measure correctly and has the potential to show us a deviation from the Standard Model induced by some new physics in a place where we might not have expected it. We heard about exciting examples of these efforts at the conference.

\subsection{Calculation of multiple weak boson production}
\label{sec:wgamma}

\begin{figure}
\centerline{
\includegraphics[width = 5 cm]{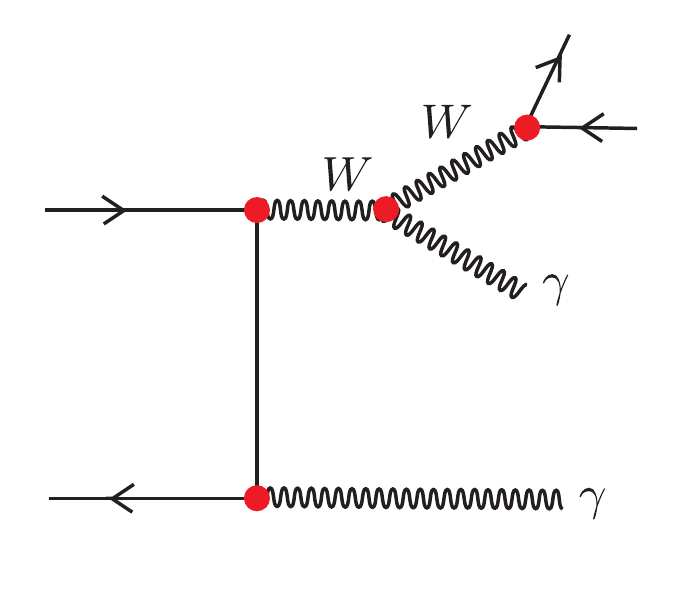}
\includegraphics[width = 5 cm]{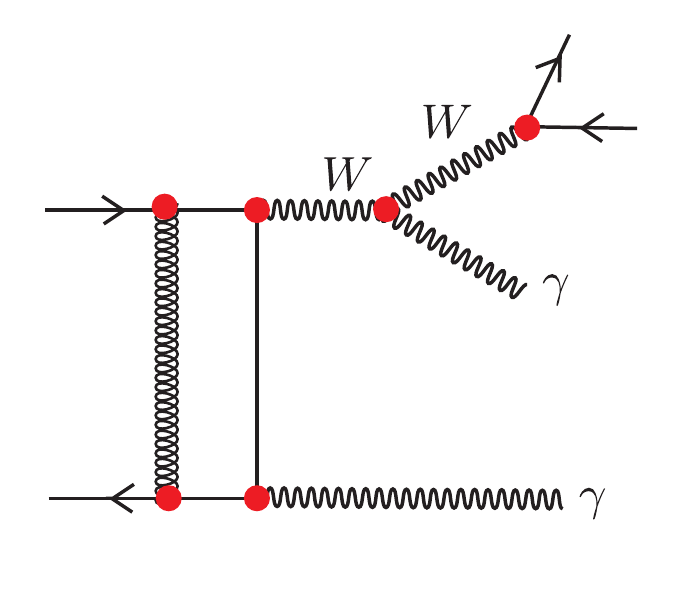}
\includegraphics[width = 5 cm]{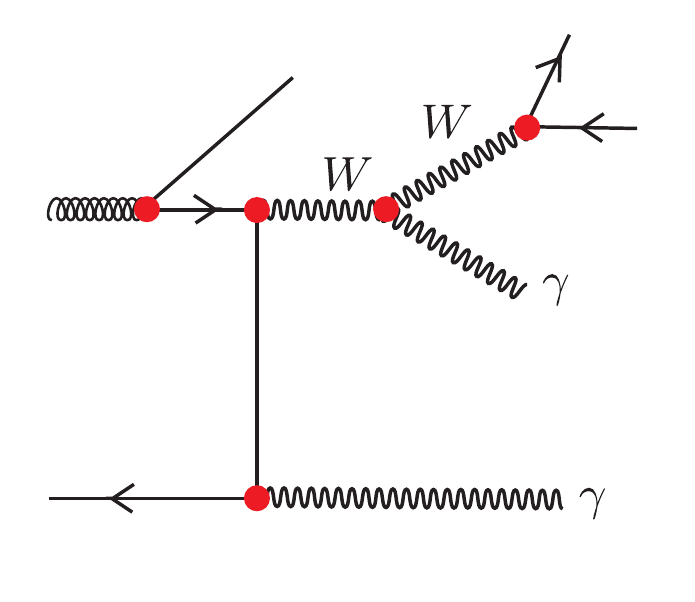}
}
\vskip -0.5 cm
\centerline{
Born graph \hskip 3 cm virtual correction \hskip 3 cm real correction
}
\caption{Graphs for production of a $W$ and two photons plus anything. 
\label{fig:W2gamma}}
\end{figure}

F.~Campanario~\cite{campanario} presented calculations for multiple electroweak boson production at next-to-leading order (NLO). His talk illustrates some general points that are worth emphasizing. Consider first the inclusive production of $W\gamma\gamma$ (that is, production of $W\gamma\gamma$ plus any number of jets). One of the the Born level diagrams is illustrated in the left hand diagram in Fig.~\ref{fig:W2gamma}. Working to NLO, there are many diagrams to add that represent virtual corrections, as in the middle diagram. Then there are also diagrams representing real corrections, that is corrections in which one more parton is emitted, as in the right hand diagram. Of course, the real diagrams and the virtual diagrams have infrared divergences, which have to be cancelled against each other and against terms arising from evolution of the incoming partons. Some of the real emission corrections introduce a new process, in which there is an initial state gluon replacing an initial state quark. There are lots of initial state gluons, so even though the NLO corrections are suppressed by a factor $\alpha_s$, they can still be large. In this case the ratio of the NLO cross section to the Born cross section (for a certain choice of scales) is as large as 3.3.

\begin{figure}
\centerline{
\includegraphics[width = 5 cm]{W2gamma2.pdf}
\includegraphics[width = 5 cm]{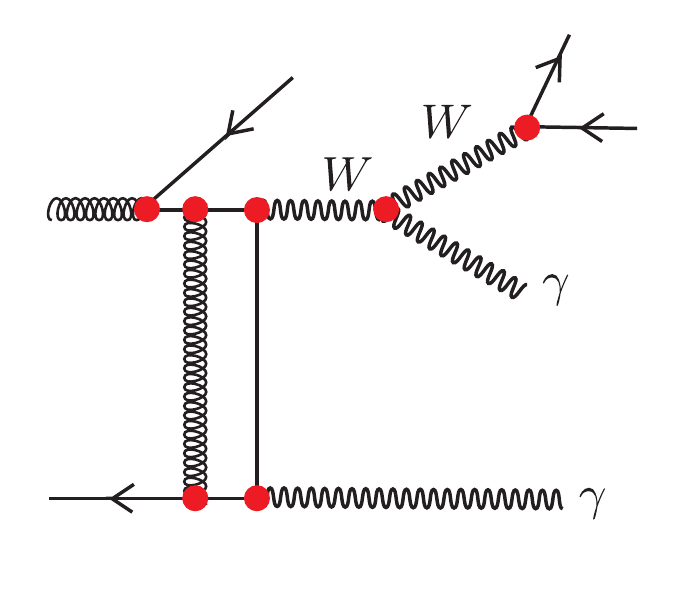}
\includegraphics[width = 5 cm]{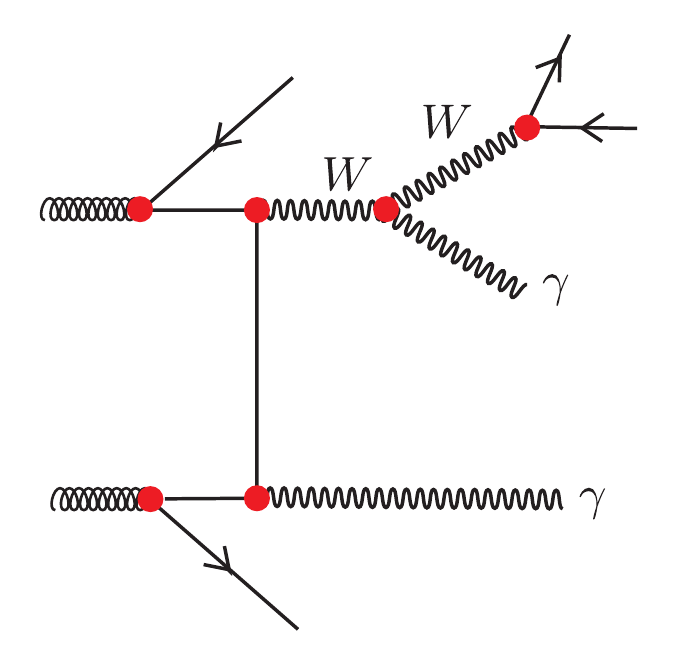}
}
\vskip -0.2 cm
\centerline{
born graph \hskip 3 cm virtual correction \hskip 3 cm real correction
}
\caption{Graphs for production of a $W$, two photons, and a jet plus anything. 
\label{fig:W2gammajet}}
\end{figure}

Unfortunately, that means one should go to NNLO. That is very difficult, but one can go part way there by considering the inclusive production of $W\gamma\gamma\ {\it jet}$. Then the previous NLO real emission graph is now one of the Born graphs, as illustrated in Fig.~\ref{fig:W2gammajet}. Working now to NLO for the inclusive production of $W\gamma\gamma\ {\it jet}$, there are again virtual corrections and real corrections, as illustrated in Fig.~\ref{fig:W2gammajet}. Again, there are new processes introduced, so the corrections need not be small. In this case the ratio of the NLO cross section to the Born cross section is as large as 1.4.

Note that this calculation contains some of the ingredients for a NNLO calculation of the inclusive production $W\gamma\gamma$, but more would be needed. As it stands, to compare the cross section for $W\gamma\gamma\ {\it jet}$ plus anything to experiment, one needs to require that the jet be measured and not have small transverse momentum.

Similar physics appeared in the talk of L.~Cieri~\cite{cieri} on inclusive diphoton production from sources other than Higgs boson decay. This is evidently of interest with respect to the Higgs searches. Here we simply omit the W-boson from the previous discussion. The same issues appear. Also, this talk emphasized the issue of a final state quark splitting into a quark plus a photon, which I did not discuss above. The photon is required to be isolated from the quark using the so-called Frixione isolation criterion (also used in the calculation presented by Campanario). In this analysis, the authors have succeeded in carrying the calculation to NNLO. That is a remarkable result.

\subsection{Progress in higher order perturbative calculations}
\label{sec:pertcalcs}

Over the past few years there has been substantial progress in performing calculations at next-to-leading order or NNLO and also in matching NLO calculations to parton showers. An example of this was visible in the experimental talk of A.~Paramonov~\cite{paramonov} on W/Z plus jets or heavy flavor production at the LHC. Paramonov showed a graph for the inclusive production plus $N$ jets for $N = 0,1,2,3$ and 4, with data compared to NLO calculations from BlackHat-Sherpa. One notes two things: first, the estimated theory error is reasonably small and, second, the agreement with experiment is within the estimated error in each case. This situation represents a substantial improvement in the theory compared to the Rencontre de Moriond of, say, ten years ago.

\begin{figure}
\centerline{
\includegraphics[width = 8 cm]{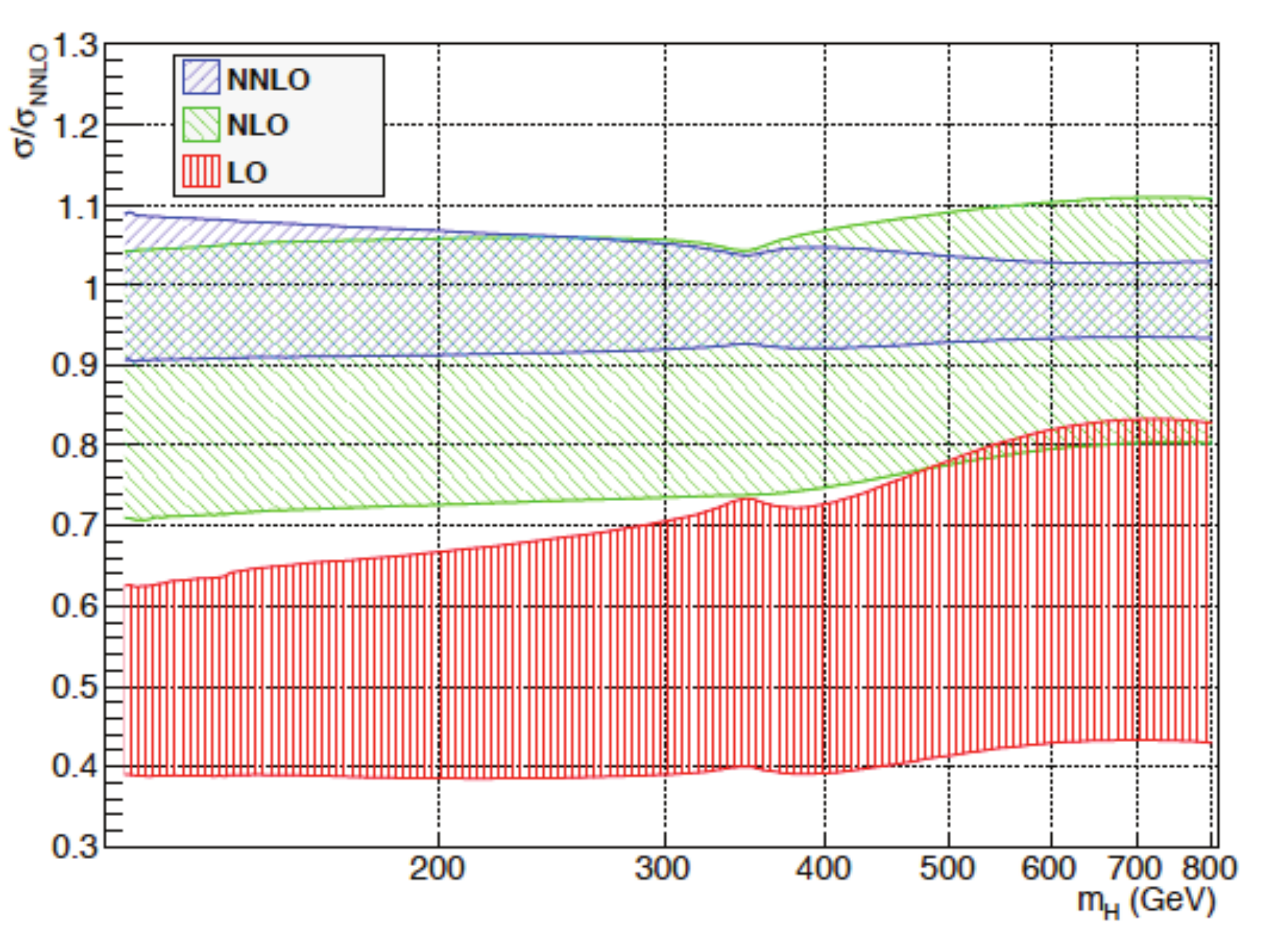}
}
\caption{Dependence of the estimated uncertainty in the Higgs production cross section on the order of perturbation theory.\protect\cite{lazopoulos}
\label{fig:higgsscale}}
\end{figure}

There were some theoretical presentations along these lines. A.~Lazopoulos~\cite{lazopoulos} showed results from a calculation of inclusive Higgs production at NNLO in QCD (in the high top mass approximation). Improvements have been added to previous results. There are also (more difficult) calculations of differential distributions. One can estimate the uncertainty in the calculation due to having left out terms of yet higher perturbative order by checking how the calculated cross section depends on the renormalization and factorization scales. As we see in Fig.~\ref{fig:higgsscale}, as we go to higher order in perturbation theory, the estimated uncertainty decreases.

Similarly, A.~Mitov~\cite{mitov} reported on progress in the calculation of the differential cross section for $ p + p \to t + \bar t + X$ at NNLO. The goal is to have a NNLO parton level event generator for this process.

\subsection{Summing large logarithms}
\label{sec:largelogs}

\begin{figure}
\centerline{
\includegraphics[width = 6 cm]{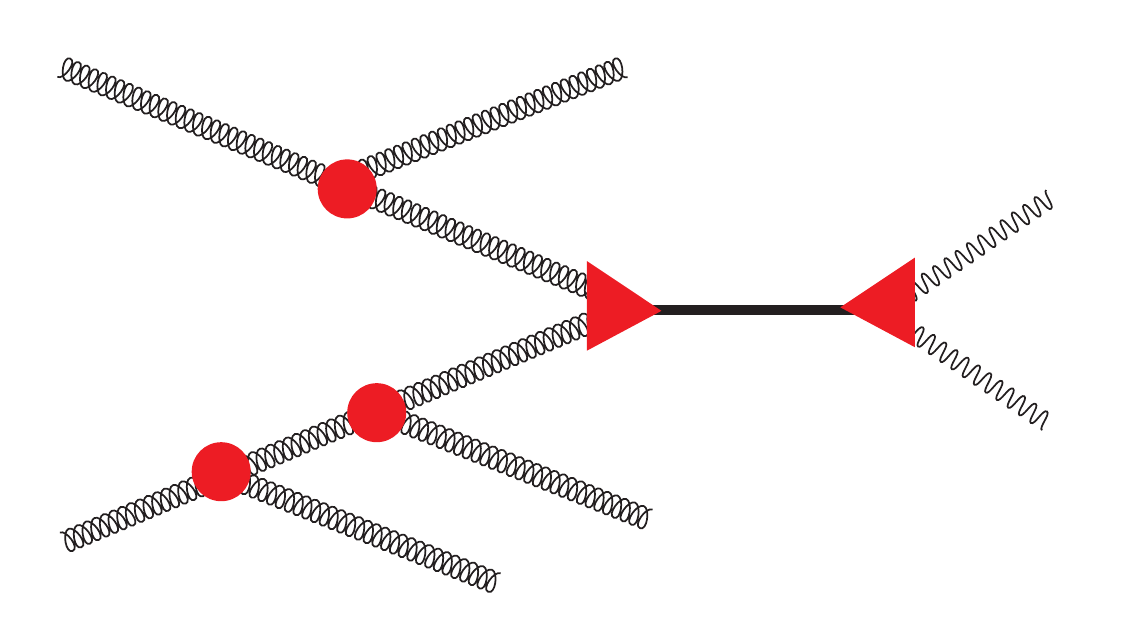}
}
\caption{The transverse momentum of a produced Higgs boson can come from multiple gluon emission from the incoming partons. The same diagram also illustrates the source of threshold logarithms.
\label{fig:higgsPT}}
\end{figure}

For a cross section that depends on two momentum scales, say $M_H$ and $p_T$, plain perturbation theory fails when $M_H^2 \gg p_T^2$ because it is an expansion in powers of $\alpha_s L^2$ where $L = \log(M_H^2/p_T^2)$. The diagrams that are responsible for the large logarithms are illustrated in Fig.~\ref{fig:higgsPT}. 

M.~Grazzini~\cite{grazzini} showed an improved calculation for the Higgs transverse momentum distribution. This includes the full kinematical information on the Higgs decay products in $H \to \gamma\gamma$, and $H \to VV$ where the vector bosons decay to leptons. It also includes matching at large $P_T^2$ to the perturbative expansion at NNLO.

G.~Ferrera~\cite{ferrera} showed an improved calculation for the transverse momentum of vector bosons, with the decay of the vector bosons included in the result.

M.~Deak~\cite{deak} discussed matching the transverse momentum dependent parton distributions used for summing small $x$ logs to a parton shower.

C.~Schwinn~\cite{schwinn} showed an improved summation of threshold logs for top pair production. If the parton distributions are falling very quickly with $x$, then emission of real gluons (as in Fig.~\ref{fig:higgsPT}) is restricted. Inside of an integration over $x$ there are large logarithms, known as threshold logarithms. For the observed cross section, the large parameter is effectively $d\log[f_{a/A}(x,\mu^2)]/dx$. For top or Higgs production at LHC, one can debate (and one did debate at the coffee breaks) the value of a summation of threshold logs compared to a full calculation at one higher order in $\alpha_s$, as in the talks of Lazopoulos~\cite{lazopoulos} and Mitov~\cite{mitov}.

\subsection{Jet cross section}
\label{sec:jets}

\begin{figure}
\centerline{
\includegraphics[width = 8 cm]{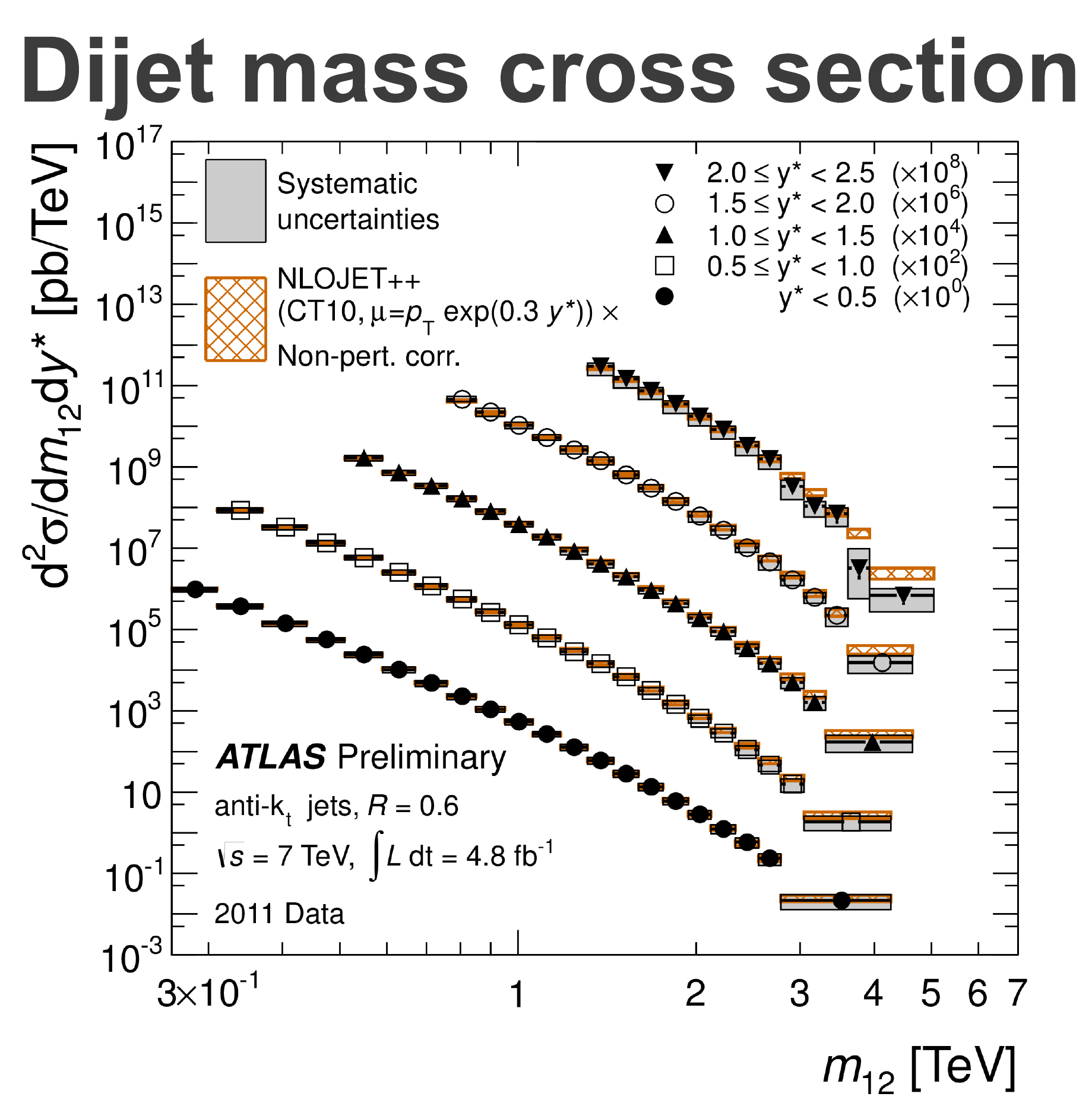}
}
\caption{Dijet mass distribution from Atlas.\protect\cite{jones}}
\label{fig:jets}
\end{figure}

There were many experimental presentations in which the data was compared to theoretical calculations. I have mentioned a couple of these, but it would not be useful to list many examples. Let me simply draw attention to one classic example that I think is exciting. G.~Jones~\cite{jones} presented jet data from the LHC. In Fig.~\ref{fig:jets}, I show the Atlas data on the dijet mass distribution in $p + p \to {\it jet} + {\it jet} + X$, grouped in bins of the rapidity difference $2y^*$ between the two jets. The data extend to large mass values, about 3 TeV. The data are compared with the NLO cross section using NLOjet++. There is some disagreement in the bin of largest $y^*$, but this is a region with some difficulties in the NLO theory. Everywhere else, there is good agreement. Most relevant is the bin of smallest $y^*$. Here is where we would see a deviation if there were a heavy object that couples to color. No deviation is seen.

\subsection{Top quark mass}
\label{sec:topmass}

We heard exciting results about direct measurements of the top quark mass. O.~Brandt~\cite{brandt} presented a measurement from the Tevatron (D0 and CDF) of $m_t = (173.2\pm 0.9) \ {\rm GeV}$, while S.~Blyweert~\cite{blyweert} presented a measurement from the CMS of $m_t = (172.6\pm 1.3) \ {\rm GeV}$. The top mass is a parameter in the QCD lagrangian that is subject to renormalization. One can adopt different prescriptions for this, among them the $\overline{\rm MS}$ prescription and the pole prescription. At the level of precision of the measurements reported, the exact definition matters. Unfortunately, the data analysis methods that give the most precise results do not precisely define what mass one is measuring. It would seem that more attention to this issue from theorists is needed.

\subsection{W boson mass}
\label{sec:Wmass}

R.~Lopez de Sa~\cite{sa} reported very precise measurements of the mass of the $W$ boson at the Tevatron: $M_W = (80387 \pm 19)\ {\rm MeV}$ for CDF and $M_W = (80376 \pm 23)\ {\rm MeV}$ for D0. He reported that at this level of precision, uncertainties in the parton distributions used in the theory are an important source of error. I asked Robert Thorne, who kindly advised me that the valence $d$ quark distribution and the $\bar u - \bar d$ distribution are mostly responsible.\cite{thorne} Perhaps LHC data can help a little to pin this down. The talk of N.~Hartland~\cite{hartland} on neural net parton distributions suggests that there may be some impact.

\section{Direct searches for new physics}
\label{sec:newphysics}

Talks at the conference showed substantial progress in looking for new physics. No definitive signals have been seen, but there is lots more to do, looking for signals that are harder to see.

G.~Altarelli~\cite{altarelli} reviewed the status many of these searches and their relation to the theoretical possibilities. (I mention some of the points he raised elsewhere in this talk.) He emphasized that the null results of searches so far puts severe constraints on the parameters of the simplest models of supersymmetry, but that there is ``plenty of room for more sophisticated versions of SUSY as a solution to the hierarchy problem.'' He noted that it is important to look for the partner particles of the third generation particles. I might suggest the motto ``start with stop.''

T.~Tait~\cite{tait} reviewed the status of searches for dark matter. Assuming that the dark matter seen in the universe from its gravitational effects consists of weakly interacting massive particles (WIMPs), the LHC has a chance to produce these particles in two ways. Either LHC proton collisions can produce the WIMPs directly, or they can produce sibling particles (perhaps squarks or gluinos) that decay into the WIMPs plus standard model particles. In either case, we look for events with missing transverse momentum. Tait explained the relation of LHC searches to non-accelerator searches: direct detection experiments that seek to discover dark matter particles from space interacting with matter on earth and indirect detection that looks, for example, for photons from dark matter annihilation in space.

M.~Spannowsky~\cite{spannowsky} summarized theoretical tools for using jet substructure to find new particles and new interactions. The idea is that a very heavy particle produced approximately at rest can decay to lighter particles that have lots of transverse momentum. In fact, in many scenarios there is a chain of successive decays. The result is high transverse momentum jets that have lots of internal structure that is characteristic of particle decays instead of normal QCD interactions. One can use this internal structure to discover the new physics. For instance, S.~Lee~\cite{lee} presented one such method. A method based on the perturbative matrix elements was presented by C.~Williams.\cite{williams}

\section{Flavor physics}
\label{sec:flavor}

Flavor physics can provide a way to look for physics beyond the Standard Model. Andreas Schopper~\cite{schopper} discussed some of the main ideas and summarized recent progress, including CKM metrology, analysis of direct CP violation, mixing induced CP violation, searches for rare decays, and a surprising finding in direct CP violation in charm.

How can flavor physics can provide a way to look for physics beyond the Standard Model?  In this talk, I view the Standard Model as a low energy effective field theory with a high cutoff energy scale. With this view, we may ask about the theory beyond the high cutoff energy scale. The parameters of the Standard Model come from the high energy theory, so the fermion mass hierarchy, the CKM matrix, and so forth are providing us clues to the high energy theory. There can be more clues. Besides the Standard Model lagrangian, we should have extra terms. For example, we might have a term
\begin{equation}
\Delta {\cal L} = 
\frac{g}{\Lambda^2} (\bar\psi \psi)^2
+ \cdots
\;\;,
\end{equation}
where $\psi$ is a quark or lepton field. The extra terms are suppressed by powers of the cutoff scale $\Lambda$. We can look for the extra terms in rare processes. 

This same analysis applies for the analysis of experiments at a few GeV scale even if $\Lambda$ is on the order of a TeV. In that case, the Standard Model is already breaking down at the LHC energy scale and this breakdown is directly accessible at the LHC, the new physics at $E >\Lambda$ is indirectly accessible via flavor physics at lower energy.

This general approach was nicely outlined by Nazila Mahmoudi~\cite{mahmoudi}. Consider $B_s$ decay. We can use
\begin{equation}
{\cal H} = - \frac{4 G_F}{\sqrt 2}\,
V_{tb} V^*_{ts} \sum_i C_i(\mu) {\cal O}_i(\mu)
\;\;.
\end{equation}
The ${\cal O}_i$ here are operators for the light partons. The coefficients $C_i(\mu)$ are calculated at $\mu = M_W$ from ${\cal L}+\Delta {\cal L}$. Then they are translated from the scale $M_W$ to the scale $m_b$ using an effective field theory in which the $W$ and $Z$ bosons are ``integrated out.'' Using this approach, there are several observables of interest. For example, one can look at the forward backward asymmetry $A_{FB}(B \to K^* \mu^+ \mu^-)$. For each observable, we need hadronic matrix elements like $\langle K^* (p+q)|{\cal O}_i | B_s(p)\rangle$. For the hadronic matrix elements, we need a calculation. Thus there are several ingredients, but the net result contains the $C_i(\mu)$, which depend on $\Delta {\cal L}$, so in the end we have a chance to learn about the physics beyond the Standard Model contained in $\Delta {\cal L}$.

\begin{figure}
\centerline{
\includegraphics[width = 10 cm]{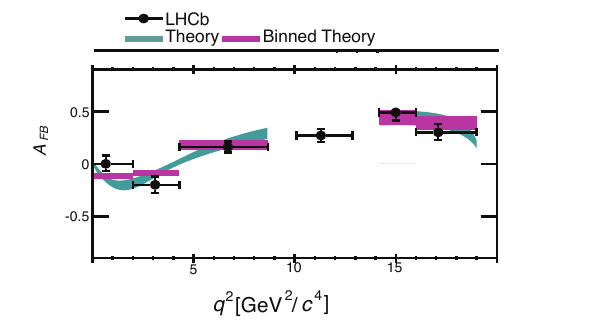}
}
\caption{Forward-backward asymmetry in $B_s$ decay.\protect\cite{parkinson}}
\label{fig:AFB}
\end{figure}

A nice example of this kind of program was presented by Cai-Dian Lu.\cite{lu} Corresponding data from LHCb was shown by Chris Parkinson.\cite{parkinson} In Fig.~\ref{fig:AFB}, I show the forward backward asymmetry $A_{FB}(B \to K^* \mu^+ \mu^-)$ as a function of the momentum transfer $q^2$. In this case, everything matches and we do not find evidence for a non-zero $\Delta {\cal L}$.

Sometimes the weakest link in this chain of argument is the assumptions that go into the calculation of the hadronic matrix elements.  Joachim Brod\cite{brod} talked the uncertainty in the calculation of the hadronic matrix elements for direct CP violation in D meson decays. Within these uncertainties, he argued how due care with modeling can allow Standard Model to plausibly explain the data. The talk of of Giulia Ricciardi~\cite{ricciardi} illustrated how uncertainties in hadronic matrix elements can be controlled. Manuel Hita-Hochgesand~\cite{hochgesand} reported a measurement by NA48/2 of two of the hadronic form factors needed for the analysis of direct CP violation in $K^\pm$ decays. In general, getting at the hadronic matrix elements needed for flavor physics is not easy. However, we have good theoretical tools: lattice gauge theory, heavy quark effective theory, soft collinear effective theory, {\it etc}.

Richard Hill~\cite{hill} reported on a possible clue to new physics that relates to the difference between electrons and muons. One can measure the proton charge radius from electron-proton scattering, or from the energy levels of the hydrogen atom, or from the energy levels of muonic hydrogen. Because a muon bound to a proton spends a lot of its time close to the proton, the measurement using the energy levels of muonic hydrogen is the most accurate. However, this measurement using muons does not agree with the two measurement methods using electrons. Hill reported that careful attention to the theory in electron-proton scattering and hydrogen energy levels does not rescue us from this discrepancy. Thus one wonders if there is some new physics that couples differently to muons and electrons. It is not so easy to see what this could be while remaining consistent with other constraints. Thus a mystery remains.

There was quite a lot of discussion about the forward backward asymmetry in top pair production at the Tevatron, which was reported by David Mietlicki.\cite{mietlicki} Produced top quarks tend to follow the proton direction while top antiquarks tend to follow the antiproton direction. Alison Lister~\cite{lister} reported analogous results from Atlas and CMS, but here measuring whether the top quark has larger $|y|$ than the antitop. While the LHC asymmetry result is inconclusive, the observed Tevatron forward backward asymmetry is larger than predicted in the Standard Model, suggesting that top quarks may couple to something not included in the Standard Model. However, some caution is needed. The asymmetry vanishes at leading order in QCD, so that the asymmetry calculated at next-to-leading order is actually calculated at the leading order at which it is nonzero. Thus one needs a calculation at yet higher perturbative order. The presentation of A.~Mitov~\cite{mitov} discussed some of these issues.

\section{Looking for the Standard Model Higgs boson}
\label{sec:higgs}

The topic on everyone's mind at this Rencontre de Moriond was the Higgs boson. Is it possible that the mechanism for electroweak symmetry breaking is the fundamental scalar field posited in the Standard Model? Let me put this question more provocatively. Is it possible that the Standard Model, with its Higgs boson, is correct as an effective field theory up to a cutoff scale $\Lambda$ that is very large compared to the Higgs field vacuum expectation value (about 250 GeV)? If so, the natural scale for the Higgs boson mass would be of order $\Lambda$ and we need ``fine tuning'' to keep it small. To stick with the Standard Model in this sense, we choose to simply ignore this problem.

\begin{figure}
\centerline{
\includegraphics[width = 8 cm]{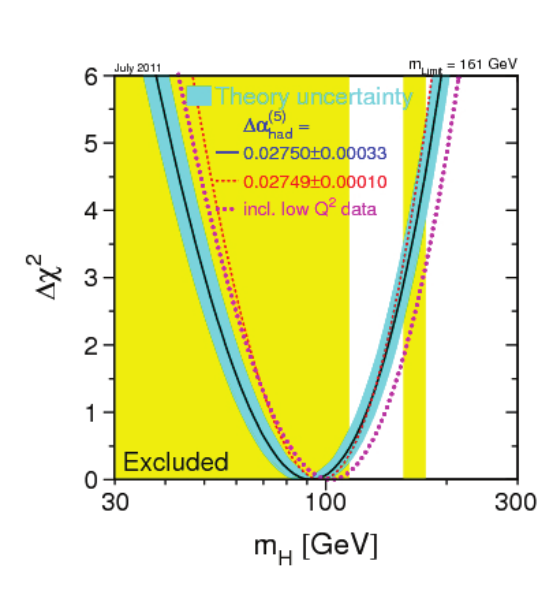}
}
\caption{Effect of electroweak precision data on the mass of the Higgs boson (from the talk of Lopes de S\'a~\protect\cite{sa}).  }
\label{fig:MH}
\end{figure}

This picture brings with it some tight constraints. First, the Higgs boson mass cannot be just anything. If I say that $\Lambda$ is at least as big as 100 TeV, then $m_h$ is bounded from below at about 100 GeV because of vacuum stability arguments and it is bounded above at about 300 GeV so as to not to produce a Landau pole into perturbation theory. This issue was discussed by G.~Altarelli.\cite{altarelli} There is more. The direct search for the Higgs boson at LEP puts a lower bound on $m_h$ at 114 GeV. Moreover, electroweak precision data puts strong constraints on the Higgs boson mass, assuming that it is the Higgs boson of the Standard Model. This is illustrated in Fig.~\ref{fig:MH}. Clearly $m_h > 200\ {\rm GeV}$ is excluded.

\begin{figure}
\centerline{
\includegraphics[width = 8 cm]{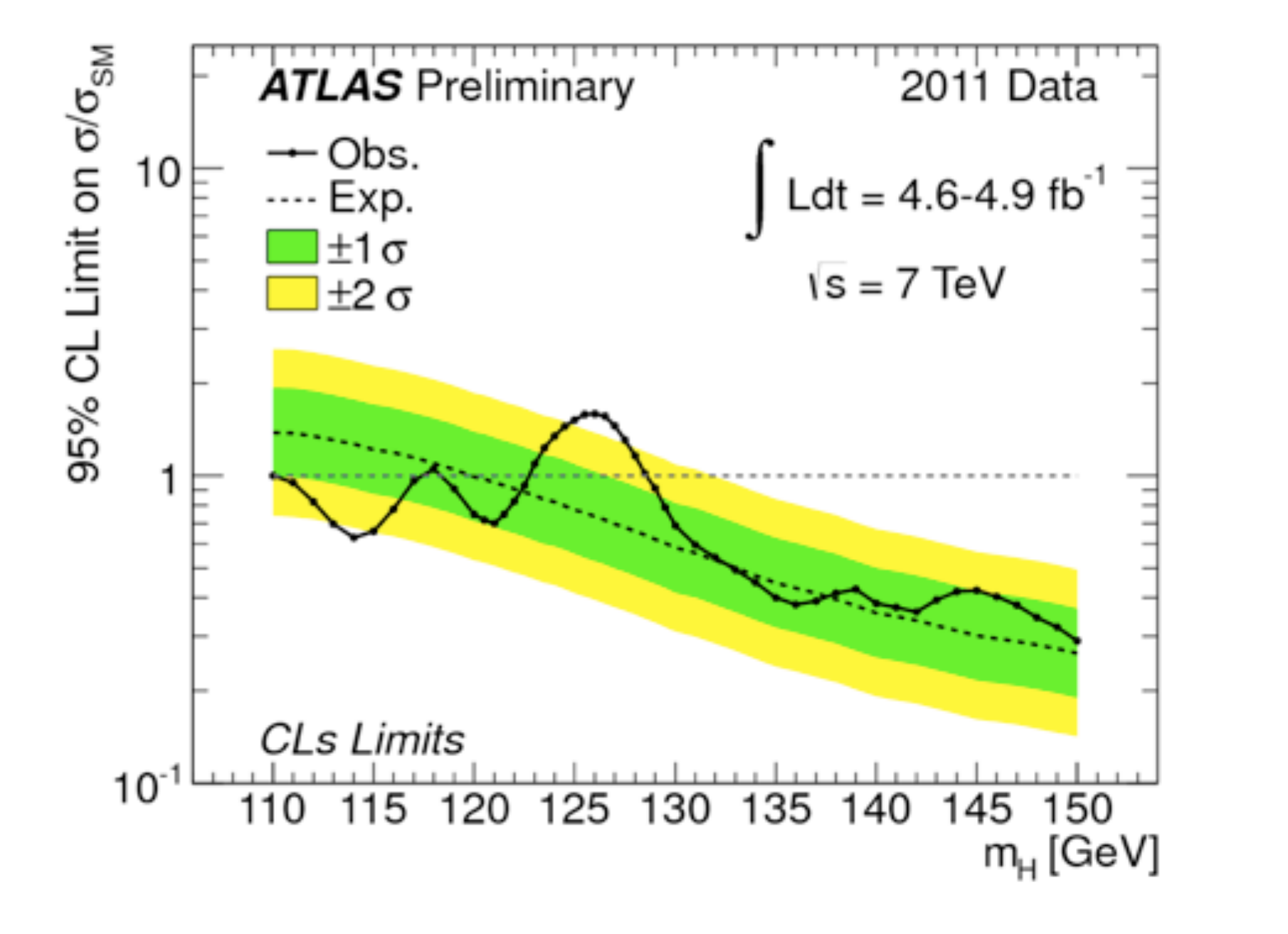}
\includegraphics[width = 7 cm]{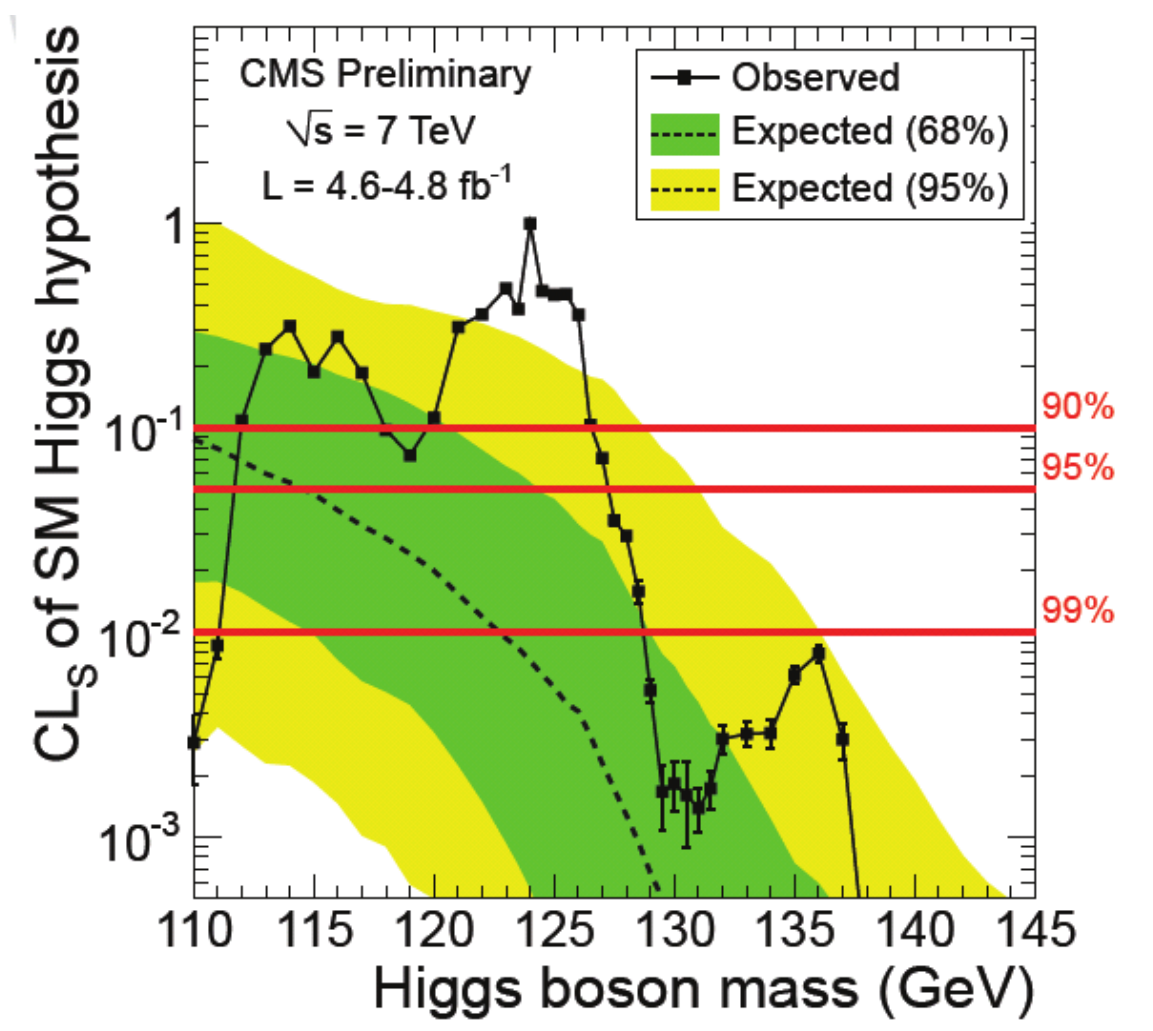}
}
\caption{Higgs boson exclusion limits from Atlas~\protect\cite{bernhard} and CMS.\protect\cite{bornheim}  }
\label{fig:higgslimits}
\end{figure}

Now the exciting news at this meeting was that the LHC experiments, with the Tevatron helping, exclude most of the available range for a Standard Model Higgs boson. A region around 125 GeV remains, as shown in Fig.~\ref{fig:higgslimits}. Furthermore, both CMS and Atlas see events that could match a SM Higgs with mass 125 GeV and that is unlikely to be a background fluctuation. See the talks of Ralf Bernhard~\cite{bernhard} and Adi Bornheim.\cite{bornheim} The Tevatron experiments also report a signal that could be $H \to b + \bar b$ with a Higgs boson mass of about 125 GeV, as reported by Daniela Bortoletto.\cite{bortoletto}

What looks like a signal for a 125 GeV Higgs boson could be a result of misestimated backgrounds and random fluctuations. It is not as convincing as the evidence for the top quark at the 1995 Rencontre de Moriond. Maybe the Standard Model Higgs boson will be ruled out with further data. If so, we will need to find a non-Standard-Model version. Then it will be significant that Atlas and CMS already can rule out a Standard-Model-like Higgs up to 540 to 600 GeV.

On the other hand, the seeming signal at around 125 GeV could well be real. Data in the 2012 LHC run will tell the story. If the signal is real, we will want to know if the found object is really the Standard Model Higgs boson. We will want to test whether there is more than one resonance seen. We will want to know whether the couplings of the resonance to W bosons, Z bosons, top quarks, bottom quarks, and tau leptons are in accordance with the Standard Model. We will want to know if the resonance really has spin zero. We will want to see if the effect of the Higgs field on W-W scattering works out as claimed in the Standard Model, with the W-W cross section not growing as the c.m. energy of the W-W system increases. Evidently, this is an ambitious program, which will not be accomplished by the end of 2012 even if the basic signal is confirmed.

This brings me back to the biblical passage that I quoted at the beginning of this talk: ``For now we see through a glass, darkly; but then face to face: now I know in part; but then shall I know \dots.'' For the Higgs boson and possible extensions of the Standard Model, the glass may be not so dark at Moriond 2013.

\section*{Acknowledgments} On behalf of, I think, all of the participants of Moriond QCD 2012, I would like to thank the organizing committee for putting this meeting together. It is a very big task to organize a meeting like this and even more of a challenge to do so while maintaining the spirit of free scientific exchange that characterizes the Rencontres de Moriond. This work was supported by the United States Department of Energy.

\end{document}